\documentclass[11pt]{article}

\begin{document}

\def\beq{\begin{equation}}
\def\eeq{\end{equation}}
\def\bea{\begin{eqnarray}}
\def\eea{\end{eqnarray}}
\def\beas{\begin{eqnarray*}}
\def\eeas{\end{eqnarray*}}
\def\ov{\overline}
\def\ot{\otimes}

\newcommand{\hf}{\mbox{$\frac{1}{2}$}}
\def\sig{\sigma}
\def\De{\Delta}
\def\af{\alpha}
\def\be{\beta}
\def\la{\lambda}
\def\ga{\gamma}
\def\ep{\epsilon}
\def\vep{\varepsilon}
\def\half{\frac{1}{2}}
\def\third{\frac{1}{3}}
\def\fth{\frac{1}{4}}
\def\sth{\frac{1}{6}}
\def\tth{\frac{1}{24}}
\def\tde{\frac{3}{2}}

\newif\ifbbB\bbBfalse                
\bbBtrue                             

\ifbbB   
 \message{If you do not have msbm (blackboard bold) fonts,}
 \message{change the option at the top of the text file.}
 \font\blackboard=msbm10 
 \font\blackboards=msbm7 \font\blackboardss=msbm5
 \newfam\black \textfont\black=\blackboard
 \scriptfont\black=\blackboards \scriptscriptfont\black=\blackboardss
 \def\Bbb#1{{\fam\black\relax#1}}
\else
 \def\Bbb{\bf}
\fi

\def\id{{1\! \! 1 }}
\def\bo{{\Bbb 1}}
\def\bI{{\Bbb I}}
\def\bC{{\Bbb C}} 
\def\bZ{{\Bbb Z}}
\def\CN{{\cal N}}

\title{{\bf Comments} on 
{\it New representations of the Hecke algebra and algebraic
Bethe Ansatz for an integrable generalized spin ladder}, 
{\small by  H.-Q. Zhou, H. Frahm and M.D. Gould, cond-mat/9911072}}
\author{{\bf Z. Maassarani} \\
\\
{\small Physics Department}\\
{\small University of Virginia}\\
{\small 382 McCormick Rd.}\\
{\small Charlottesville, VA,  
22903 USA}\thanks{Email address: zm4v@virginia.edu} \\}
\date{}
\maketitle

\begin{abstract}
The authors of cond-mat/9911072  claim to introduce 
``new representations of the Hecke algebra.''
These representations are shown to be
the XXC models introduced two years ago in solv-int/9712008, 
and repeatedly studied and referred  to in subsequent papers. 
They are in fact representations of the Temperley-Lieb algebra. 
Several other remarks are  made and mistakes are pointed out. 
\end{abstract}
\vspace*{2.5cm}

{\bf 1)} The authors of   \cite{zfg} claim to 
have recently constructed a ``{\it novel class of representations of the 
Hecke algebra.}'' It is shown here that these representations 
were already  known. They were discovered {\bf two years} ago and named 
XXC models \cite{xxc}. 

Start from the matrix  $\check{R}(x)$ defined by (2) and (4) in \cite{zfg}:
\beq
\check{R} (x) = \check{R} -x q^{2}\check{R}^{-1},
\label{rcx}
\eeq
where 
\bea
\check R &=&  q \sum _{\af \in A,\be \in B}
  \left(X^{\af \be}\otimes X^{\be \af}
   +X^{\be \af}\otimes X^{\af \be}\right)
+ \left(q^2+1\right)
         \left(\frac{1}{2}-\sum_{\af\in A} X^{\af\af}\right)\otimes
         \left(\frac{1}{2}-\sum_{\af\in A} X^{\af\af}\right)  
\nonumber\\ 
& &+ \frac {q^2-1}{2}
  \sum _{\af \in A}\left(X^{\af \af}\otimes \bI- \bI\otimes X^{\af \af}\right)
  +\left(-\frac {1}{4} q^2+\frac {3}{4}\right) \bI\otimes\bI \label{rc}
\eea
and the index sets $A$ and $B$ take values from $\{
1,2,\dots,m\} $ and $\{ m+1,m+2,\dots,n\}$, respectively.  
The $n\times n$ matrix $X^{\af \be}$ has only 
one non-vanishing element, a 1 at row $a$ and column $b$.
The matrix $X^{\af\be}$ is denoted by $E^{\af\be}$ below.

Now consider the operators defined in \cite{xxc}.
Let $n$, $n_1$ and $n_2$ be three positive integers such that
$n_1+n_2=n$, 
and $A$, $B$ be two disjoint sets whose union is the set of basis 
states of $\bC^n$, with card$(A)=n_1$ and card$(B)=n_2$. 
Let also
\bea
P^{(3)}&=&\sum_{a\in A}\sum_{\beta\in B}\left( E^{\beta a}\otimes
E^{a\beta} +  E^{a\beta}\otimes E^{\beta a}\right)\\
P^{(1)}&=&P^{(+)}+P^{(-)}=
\sum_{a\in A}\sum_{\beta\in B}\left( E^{a a}\otimes E^{\beta\beta}+ 
E^{\beta\beta}\otimes E^{a a}\right)\\
P^{(2)}&=&\sum_{a,a^{'}\in A} E^{a a}\otimes
E^{a^{'} a^{'}} + \sum_{\beta,\beta^{'}
\in B} E^{\beta\beta}\otimes E^{\beta^{'} \beta^{'}}
\eea
The $n_1 n_2$ parameters $x_{a\beta}$ of \cite{xxc} 
were taken all equal to one.  
Latin indices belong to $A$ while Greek indices belong to $B$.
 
With obvious identifications and  $m\rightarrow n_1$ and  $n_2=n-m$,  
a simple and short calculation allows one to find $\check{R}^{-1}$,
and to  rewrite (\ref{rcx}) as follows:
\beq
\check{R} (x) = q(1-x) P^{(3)} + (1-q^2)(x P^{(+)}+  P^{(-)}) +
(1-x q^2)P^{(2)} \label{rxxc}
\eeq
With the usual passage to additive variable, one obtains 
the XXC models in their asymmetric form {\it i.e.\/} their Hecke algebra
form \cite{amp}. 
A gauge transformation (a special type of similarity  transformation)
gives  the `deformed free-fermion' and  Temperley-Lieb forms \cite{xxc,amp}.
\\

{\bf 2)} The XXC models are not just representations of Hecke algebra,
they are {\bf more accurately} representations of the Temperley-Lieb
algebra \cite{xxc}. They have  an underlying $sl(2)$ structure.  
The natural generalizations to $sl(m+1)$ are  the multiplicity $A_m$ models, which include the XXC models \cite{amp}. They are representations of the
Hecke algebra. 
\\

{\bf 3)} The matrix  (\ref{rcx}),  at $x=1$, 
is more accurately equal to $(1-q^2) \bI$,  and not $\bI$. 
\\

{\bf 4)} The Hamiltonian (5) in \cite{zfg} is  exactly the XXC Hamiltonian
(18) in \cite{xxc}. (The boundary terms do not contribute under periodic
boundary conditions.) See also (22) in \cite{amp}.
The $J$-term was added by hand, 
and commutes with all the conserved quantities of the  model \cite{xxc}. 
\\

{\bf 5)} In equation (13) of \cite{zfg}, $\Lambda^{(1)}$ is independent
of its arguments.  After equation (13): ``{\it Unfortunately,
it} ($\tau ^{(1)} = tr_* [P_{1*} \cdots P_{M*}]$) 
{\it can not be diagonalized directly in terms of the algebraic Bethe
Ansatz.}'' Rather {\bf fortunately}, 
any such unit-shift operator is {\bf automatically}
diagonalized by algebraic Bethe Ansatz (for periodic boundary conditions)
every time $\check{R}(1)$ 
is proportional to the identity operator ({\it i.e.} $\check{R}$ is regular). 
This is one of the most basic aspects of the algebraic Bethe Ansatz in 
the framework of the QISM. It is enough to  consider {\bf any}
regular $\check{R}$-matrix with 3 states per site. Examples include
the spin-1 $sl(2)$ matrix and  
the $sl(3)$ matrix,  both trigonometric or rational.  
\\

{\bf 6)} {\it Footnote 1:} The thermodynamics may be affected by the increased
degeneracy of all the states. The phases of certain roots of unity
with order proportional to the number of sites, may affect the finite-size
corrections. All this requires careful checks. 
\\

{\bf 7)} The ``mapping'' of \cite{zfg} is misleading and meaningless.
Most Bethe Ansatz equations look alike and  are  determined  by the
Dynkin diagram of the Lie algebra and weight of the representation at hand. 
It necessary to complement them with the eigenvalues and eigenvectors
to which they refer, and to take into account the degeneracies and the
specific features of the model. 
The mapping to the Perk-Shultz models is false. 

Case in point: The Bethe Ansatz equations (16) in \cite{zfg} are simply wrong.


\begin{thebibliography}{30}

\bibitem{zfg} H.-Q. Zhou, H. Frahm and M.D. Gould, 
{\it New representations of the Hecke algebra and algebraic
Bethe Ansatz for an integrable generalized spin ladder}, cond-mat/9911072.
\bibitem{xxc} Z.~Maassarani, Phys. Lett. A  {\bf 244} (1998) 160--164,
solv-int/9712008. 
\bibitem{amp} Z.~Maassarani, Eur. Phys. J. B {\bf 7} (1999) 627--633,
solv-int/9805009, Erratum {\bf 9} (1999) 371. 

\end{thebibliography}
\end{document}